\documentstyle[prl,aps]{revtex}
\begin{document}
\twocolumn[\hsize\textwidth\columnwidth\hsize
           \csname @twocolumnfalse\endcsname
\title{Nonlocal corrections to the Boltzmann equation for dense Fermi
systems}
\author{V\'aclav \v Spi\v cka and Pavel Lipavsk\'y}
\address{Institute of Physics, Academy of Sciences, Cukrovarnick\'a 10,
16200 Praha 6, Czech Republic}
\author{Klaus Morawetz}
\address{Fachbereich Physik, University Rostock, D-18055 Rostock,
Germany}
\maketitle
\begin{abstract}
A kinetic equation which combines the quasiparticle drift of Landau's
equation with a dissipation governed by a nonlocal and noninstant
scattering integral in the spirit of Snider's equation for gases is
derived. Consequent balance equations for the density, momentum and
energy include quasiparticle contributions and the second order quantum
virial corrections. The medium effects on binary collisions are shown to
mediate the latent heat, i.e., an energy conversion between correlation
and thermal energy. An implementation to heavy ion collisions is
discussed.
\end{abstract}
\vskip2pc]
\section{Introduction}
One of the goals of the recent nuclear physics is to find the equation
of state of nuclear matter. Indeed, the dependence of the pressure on
the density of nucleons is a crucial input for a hydrodynamical modeling
of heavy ion collisions or of astrophysical events like the big bang,
supernova explosions and neutron stars \cite{SG86}.

In the absence of any direct measurement, it is hoped that the equation
of state can be deduced from heavy ion collisions via the following
scheme. Heavy ion collision data are fitted with the Boltzmann equation
(BE)
\begin{eqnarray}
&&{\partial f_1\over\partial t}+{\partial\varepsilon_1\over\partial k}
{\partial f_1\over\partial r}-{\partial\varepsilon_1\over\partial r}
{\partial f_1\over\partial k}
\nonumber\\
&&=\sum_b\int{dpdq\over(2\pi)^5}
\delta\left(\varepsilon_1+\varepsilon_2-
\varepsilon_3-\varepsilon_4\right)
\nonumber\\
&&\times |T|^2\left(\varepsilon_1+\varepsilon_2,k,p,q,t,r\right)
\nonumber\\
&&\times
\Bigl[f_3f_4\bigl(1-f_1\bigr)\bigl(1-f_2\bigr)-
\bigl(1-f_3\bigr)\bigl(1-f_4\bigr)f_1f_2\Bigr].
\label{1}
\end{eqnarray}
Arguments of distributions $f$ and energies $\varepsilon$ are shortened
as $f_1\equiv f_a(k,r,t)$, $f_2\equiv f_b(p,r,t)$,
$f_3\equiv f_a(k-q,r,t)$, and $f_4\equiv f_b(p+q,r,t)$, with momenta
$k,p,q$, coordinate $r$, time $t$, and spin and isospin $a,b$. Once the
differential cross sections $|T|^2$ and the functional dependence of
energy $\varepsilon$ on the distribution $f$ are fitted, the equation of
state is evaluated from the kinetic equation.

This scheme has two drawbacks. First, accessible fits of the
quasiparticle energy $\varepsilon$ are not sufficiently reliable since
two possible fits, momentum-dependent and momentum-independent, result
in very contradictory predictions giving hard and soft equations of
state, respectively\cite{BG88}. When this more or less technical problem
is resolved in future, one has to face the second drawback: the BE
is not thermodynamically consistent with virial corrections to the
equation of state. This problem is principal for ``how can one infer the
equation of state from the BE if the two equations are not
consistent?''. A consistency between the kinetic and the thermodynamic
theories is a general question for the quantum statistics exceeding the
merits of the nuclear matter. Here we approach this question from
nonequilibrium Green's functions. It is shown that the consistency is
achieved by a consistent treatment of the quasiclassical limit which
results in nonlocal and noninstant corrections to the scattering
integral of the BE.

The need of nonlocal corrections can be seen on the classical gas of
hard spheres. In the scattering integral of (\ref{1}), all space
arguments of the distributions are identical, i.e., colliding particles
$a$ and $b$ are at the same space point $r$. In reality, these particles
are displaced by the sum of their radii. This inconsistency has been
noticed by Enskog \cite{CC90} and cured by nonlocal corrections to the
scattering integral. The equation of state evaluated from the kinetic
equation with the nonlocal scattering integral is of the van der Waals
type covering the excluded volume \cite{CC90,HCB64}. For nuclear matter,
Enskog's corrections has been first discussed by Malfliet \cite{M84}
and recently implemented by Kortemayer, Daffin and Bauer \cite{KDB96}.

The noninstant corrections are closer to the chemical picture of
reacting gases. In the scattering integral of (\ref{1}), all time
arguments of the distributions are identical what implies that the
collision is instant. In reality, the collision has a finite duration
which might be quite long when two particles form a resonant state.
The resonant two-particle state behaves as an effective short-living
molecule. Like in reacting gases \cite{HCB64}, the presence of these
molecules reduces the pressure since it reduces the number of freely
flying particles. The finite duration of nucleon-nucleon collisions and
its thermodynamic consequences has been for the first time discussed
only recently by Danielewicz and Pratt \cite{DP96}. The noninstant
scattering integral and its consequencies for the linear response has
been also discussed for electrons in semiconductors scattered by
resonant levels \cite{SLM97}.

Except for dense Fermi systems, the above intuitively formulated
nonlocal and noninstant corrections has been confirmed by systematic
approaches. For classical gases, this theory was developed already by
Bogoliubov and Green \cite{B46,G52}. Obtained gradient contributions to
the scattering integral are the lowest order terms of the virial
expansion in the kinetic equation \cite{comdiv}. The first quantum
kinetic equation with nonlocal corrections has been derived by Snider
\cite{S60}. Recently, it has been recognized that Snider's equation is
not consistent with the second order virial corrections to equations of
state. A consistent quantum mechanical theory of the virial corrections
to the BE has been developed from the multiple scattering expansion in
terms of M\o ller operators \cite{NTL91} and confirmed by Balescu's
formalism \cite{H90}.

Presented treatment extends the nonlocal and noninstant corrections to
dense Fermi systems. We follow Baerwinkel \cite{B69} in starting from
nonequilibrium Green's functions and keeping all gradient contributions
to the scattering integral. Baerwinkel's results are limited to low
densities (to avoid medium effects on binary collisions) and not
consistent (since he uses the quasiparticle approximation). Here we
describe the binary collisions by the Bethe-Goldstone T-matrix which
includes the medium effects. Instead of the quasiparticle approximation,
the {\em extended} quasiparticle approximation is used. This extension
is sufficient to gain consistency of the kinetic theory with the virial
corrections to thermodynamic quantities.

\section{Extended quasiparticle picture}
We start our derivation of the kinetic equation from the quasiparticle
transport equation first obtained by Kadanoff and Baym \cite{D84,SL95}
\begin{equation}
{\partial f_1\over\partial t}+{\partial\varepsilon_1\over\partial k}
{\partial f_1\over\partial r}-{\partial\varepsilon_1\over\partial r}
{\partial f_1\over\partial k}=
z_1(1-f_1)\Sigma^<_{1,\varepsilon_1}-z_1f_1\Sigma^>_{1,\varepsilon_1}.
\label{2}
\end{equation}
Like in (\ref{1}), quasiparticle distribution $f$, quasiparticle energy
$\varepsilon$ and wave-function renormalization $z$ are functions of
time $t$, coordinate $r$, momentum $k$ and spin and isospin $a$.
Self-energy $\Sigma$, taken from nonequilibrium Green's function in the
notation of Kadanoff and Baym \cite{D84}, is moreover a function of
energy $\omega$, however, it enters the transport equation only by its
value at pole $\omega=\varepsilon_1$.

Particular forms of the quasiparticle energy and the scattering integral
we derive for a model and an approximation used in nuclear matter for
heavy ion collisions in the non-relativistic energy domain. The system
is composed of protons and neutrons of equal mass $m$. They interact via
an instant potential $V$. We assume no spin-flipping mechanism. As
common, the self-energy is constructed from the two-particle T-matrix
$T^R$ in the Bethe-Goldstone approximation \cite{D84,MR94} as
[$T^R_{\rm sc}\!(1,2,3,4)\!=\!(1\!-\!\delta_{a_1a_2})T^R\!(1,2,3,4)\!+\!
{1\over\sqrt{2}}\delta_{a_1a_2}(T^R\!(1,2,3,4)\!-\!T^R\!(1,2,4,3))$]
\begin{eqnarray}
\Sigma^<(1,2)&=&
T^R_{\rm sc}(1,\bar 3;\bar 5,\bar 6)T^A_{\rm sc}(\bar 7,\bar 8;2,\bar 4)
\nonumber\\
&\times &G^>(\bar 4,\bar 3)G^<(\bar 5,\bar 7)G^<(\bar 6,\bar 8),
\label{3}
\end{eqnarray}
and $\Sigma^>$ is obtained from (\ref{3}) by an interchange
$>\leftrightarrow <$. Here, $G$'s are single-particle Green's functions,
numbers are cumulative variables, $1\equiv (t_1,r_1,a_1)$, and bars
denote internal variables that are integrated over. Before (\ref{3}) is
plugged in (\ref{2}), it has to be transformed into the mixed
representation, [off-diagonal elements in spin and isospin are excluded,
$a_1=a_2=a$]
\begin{eqnarray}
\Sigma^<(1,2)&=&\int{d\omega\over 2\pi}{dk\over(2\pi)^3}
{\rm e}^{ik(r_1-r_2)-i\omega(t_1-t_2)}
\nonumber\\
&\times &\Sigma^<_a\left(\omega,k,r,t\right)_
{r={r_1+r_2\over 2},t={t_1+t_2\over 2}},
\label{4}
\end{eqnarray}
and all Green's functions in (\ref{3}), too.

The self-energy $\Sigma$ is a functional of Green's functions $G$. This
functional $\Sigma[G]$ is converted to the functional of the
quasiparticle distribution $\Sigma_\varepsilon[f]$ via the extended
quasiparticle approximation \cite{SL95,BKKS96}
[$z_1=1+\left.
{\partial\over\partial\omega}{\rm Re}\Sigma_{1\omega}\right|_
{\varepsilon_1}$]
\begin{equation}
G^{\begin{array}{c}>\\[-2mm] <\end{array}}_{1,\omega}=
\left(\!\begin{array}{c}1\!-\!f_1\\ f_1\end{array}\!\right)
2\pi z_1\delta(\omega-\varepsilon_1)+{\rm Re}{\Sigma^{\begin{array}{c}
>\\[-2mm] <\end{array}}_{1,\omega}\over(\omega-\varepsilon_1)^2},
\label{5}
\end{equation}
where $G_{1,\omega}\equiv G_a(\omega,k,r,t)$ and similarly $\Sigma$.
Unlike the plain quasiparticle approximation (without the second term)
used by Baerwinkel \cite{B69}, approximation (\ref{5}) leads to the
consistent theory. The first term brings the on-shell quasiparticle
part, the second term is the off-shell contribution.

The off-shell part plays four-fold role. First, it justifies the kinetic
equation (\ref{2}). Equation (\ref{2}) has been originally derived from
the plain quasiparticle approximation neglecting the off-shell
drift.\footnote{The well known term $[{\rm Re}G,\Sigma^<]$.} The
off-shell part
of $G^<$ in (\ref{5}) compensates the off-shell drift so that (\ref{2})
is recovered without uncontrollable neglects \cite{SL95}. Second, in the
quasiparticle energy $\varepsilon_1={k^2\over 2m_a}+{\rm Re}\Sigma^R_
{1,\varepsilon_1}$, the off-shell part brings contributions that are
essential for the correct binding energy \cite{KM93}. Third, (\ref{5})
provides Wigner's distribution $\rho=\int{d\omega\over 2\pi}G^<$ as a
functional of the quasiparticle distribution $f$ \cite{SL95,KM93}.
Fourth, in the scattering integral of (\ref{2}), the off-shell part
results in sequential three-particle processes with the off-shell
propagation between the two composing binary processes. Since the
three-particle processes are beyond the scope of the present paper,
they are excluded from scattering integral.

\section{Non-local scattering integral}
Now the approximation is specified and we can start to simplify the
scattering integral. In contrast to previous treatments of degenerated
systems, we keep all terms linear in gradients. The gradient expansion
of the self-energy (\ref{3}) is a straightforward but tedious task. It
results in a one nongradient and nineteen gradient terms that are
analogous to those found within the chemical physics \cite{NTL91,H90}.
All these terms can be recollected into a nonlocal and noninstant
scattering integral that has an intuitively appealing structure of the
scattering integral in the BE (\ref{1}) with Enskog-type shifts of
arguments.\footnote{The basic idea of the recollection can be
demonstrated on the following rearrangement of the gradient
approximation of a matrix product $C(1,2)=A(1,\bar 3)B(\bar 3,2)$. In
the mixed representation
$C=AB+{i\over 2}\left(
{\partial A\over\partial\omega}{\partial B\over\partial t}-
{\partial A\over\partial t}{\partial B\over\partial\omega}-
{\partial A\over\partial k}{\partial B\over\partial r}+
{\partial A\over\partial r}{\partial B\over\partial k}\right)$,
see \protect\cite{D84,SL95}. We denote $\varphi={i\over 2}\ln A$ and
rearrange the product as $C=A\left(B+
{\partial \varphi\over\partial\omega}{\partial B\over\partial t}-
{\partial \varphi\over\partial t}{\partial B\over\partial\omega}-
{\partial \varphi\over\partial k}{\partial B\over\partial r}+
{\partial \varphi\over\partial r}{\partial B\over\partial k}\right)$.
The gradient term in brackets can be viewed as a linear expansion of $B$
with all arguments shifted as \mbox{$C=A(\omega,k,r,t)B
\left(\omega\!-\!{\partial \varphi\over\partial t},
k\!+\!{\partial \varphi\over\partial r},
r\!-\!{\partial \varphi\over\partial k},
t\!+\!{\partial \varphi\over\partial\omega}\right)$}.}
In agreement with \cite{NTL91,H90}, all gradient corrections result
proportional to derivatives of the scattering phase shift
\mbox{$\phi={\rm Im\ ln}T^R_{\rm sc}(\Omega,k,p,q,t,r)$},
\begin{equation}
\begin{array}{lclrcl}\Delta_t&=&{\displaystyle
\left.{\partial\phi\over\partial\Omega}
\right|_{\varepsilon_1+\varepsilon_2}}&\ \ \Delta_2&=&
{\displaystyle\left({\partial\phi\over\partial p}-
{\partial\phi\over\partial q}-{\partial\phi\over\partial k}
\right)_{\varepsilon_1+\varepsilon_2}}\\ &&&&&\\ \Delta_E&=&
{\displaystyle\left.-{1\over 2}{\partial\phi\over\partial t}
\right|_{\varepsilon_1+\varepsilon_2}}&\Delta_3&=&
{\displaystyle\left.-{\partial\phi\over\partial k}
\right|_{\varepsilon_1+\varepsilon_2}}\\ &&&&&\\ \Delta_K&=&
{\displaystyle\left.{1\over 2}{\partial\phi\over\partial r}
\right|_{\varepsilon_1+\varepsilon_2}}&\Delta_4&=&
{\displaystyle-\left({\partial\phi\over\partial k}+
{\partial\phi\over\partial q}\right)_{\varepsilon_1+\varepsilon_2}}.
\end{array}
\label{8}
\end{equation}
After derivatives, $\Delta$'s are evaluated at the energy shell
$\Omega\to\varepsilon_1+\varepsilon_2$. The corrected BE with the
collected gradient terms then reads
[$\Delta_r={1\over 4}(\Delta_2+\Delta_3+\Delta_4)$]
\begin{eqnarray}
&&{\partial f_1\over\partial t}+{\partial\varepsilon_1\over\partial k}
{\partial f_1\over\partial r}-{\partial\varepsilon_1\over\partial r}
{\partial f_1\over\partial k}
\nonumber\\
&&=\sum_b\int{dpdq\over(2\pi)^5}\delta\left(\varepsilon_1+\varepsilon_2-
\varepsilon_3-\varepsilon_4+2\Delta_E\right)\nonumber\\
&&\times z_1z_2z_3z_4
\Biggl(1-{1\over 2}{\partial\Delta_2\over\partial r}
-{\partial\bar\varepsilon_2\over\partial r}
{\partial\Delta_2\over\partial\omega}\Biggr)
\nonumber\\
&&\times
|T_{\rm sc}^R|^2\!\left(\varepsilon_1\!+\!\varepsilon_2\!-\!
\Delta_E,k\!-\!{\Delta_K\over 2},p\!-\!{\Delta_K\over 2},
q,r\!-\!\Delta_r,t\!-\!{\Delta_t\over 2}\!\right)
\nonumber\\
&&\times\Bigl[f_3f_4\bigl(1-f_1\bigr)\bigl(1-f_2\bigr)-
\bigl(1-f_3\bigr)\bigl(1-f_4\bigr)f_1f_2\Bigr].
\label{9}
\end{eqnarray}
Unlike in (\ref{1}), the subscripts denote shifted arguments:
$f_1\equiv f_a(k,r,t)$, $f_2\equiv f_b(p,r\!-\!\Delta_2,t)$,
$f_3\equiv f_a(k\!-\!q\!-\!\Delta_K,r\!-\!\Delta_3,t\!-\!\Delta_t)$, and
$f_4\equiv f_b(p\!+\!q\!-\!\Delta_K,r\!-\!\Delta_4,t\!-\!\Delta_t)$.

The $\Delta$'s are effective shifts and they represent mean values of
various nonlocalities of the scattering integral. These shifts enter
the scattering integral in the form known from the theory of gases
\cite{CC90,NTL91,H90}, however, the set of shifts is larger due to the
medium effects on the binary collision that are dominated by the Pauli
blocking of the internal states of the collision.

The physical meaning of the $\Delta$'s is best seen on gradually more
complex limiting cases:\\[2mm]
(o) Sending all $\Delta$'s to zero, (\ref{9}) reduces to the BE
(\ref{1}).\\[2mm]
(i) In the classical limit for hard spheres of the diameter $d$, the
scattering phase shift $\phi\to\pi-|q|d$ gives $\Delta_4=\Delta_2=
{q\over|q|}d$ and all other $\Delta$'s are zero. The Enskog's nonlocal
corrections are thus recovered.\\[2mm]
(ii) For a collision of two isolated particles interacting via the
single-channel separable potential, the scattering phase
shift does not depend on $r$, $t$, $q$ and $k-p$, while it depends on
$k+p$ exclusively via the energy dependency, $\phi\to\phi(\Omega-
{1\over 4m}(k+p)^2)$. Then $\Delta_{E,K,2}=0$ and $\Delta_{3,4}=
{k+p\over 2m}\Delta_t$. Since ${k+p\over 2m}$ is the center-of-mass
velocity, the displacements $\Delta_4=\Delta_3$ represent a distance
over which particles fly together as a molecule.\\[2mm]
(iii) For two isolated particles interacting via a general spherical
potential, the scattering phase shift reflects the translational
and the spherical symmetries. From translation of the center of mass
during $\Delta_t$ follows the relation for the molecular flight
${1\over 2}(\Delta_4+\Delta_3-\Delta_2)={k+p\over 2m}\Delta_t$. From the
spherical symmetry follows that the sum of relative coordinates of the
particles $a$ and $b$ at the end and at the beginning of collision has
the direction of the transferred momentum
${1\over 2}(\Delta_4-\Delta_3+\Delta_2)={q\over|q|}d$ (Enskog-type
shift), and that the difference has the perpendicular in-plane direction
${1\over 2}(\Delta_4-\Delta_3-\Delta_2)={k-p-q\over|k-p-q|}\alpha$
(rotation of the molecule). The nonlocality of the collision is thus
given by three scalars $\Delta_t$, $d$ and $\alpha$.\\[2mm]
(iv) For effectively isolated two-particle collisions (no Pauli blocking
but mean-field contributions $U_a(r,t)$ to the energy, i.e., $\phi\to
\phi(\Omega\!-\!{1\over 4m}(k\!+\!p)^2\!-\!U_a\!-\!U_b,\ldots)$) all
$\Delta$'s become nonzero. The space displacements are the same as in
(iii). The energy and the momentum corrections
$2\Delta_{E,K}=-\Delta_t{\partial\over\partial t,r}(U_a+U_b)$,
represent the energy and the momentum which the effective molecule gains
during its short life time $\Delta_t$. These corrections are in fact of
three-particle nature, however, only on the mean-field level, thus
formally within the binary process. The three scalars, $\Delta_t$, $d$
and $\alpha$, are still sufficient to parameterize all $\Delta$'s. We
note that in the energy-conserving $\delta$ function all $\Delta$'s and
mean-fields compensate
$\delta\left(\varepsilon_1\!+\!\varepsilon_2\!-\!\varepsilon_3\!-\!
\varepsilon_4\!-\!2\Delta_E\right)\!=\!\delta\left({k^2\over 2m}\!+\!
{p^2\over 2m}\!-\!{(k-q)^2\over 2m}\!-\!{(p+q)^2\over 2m}\right)$. This
limit describes the dilute quantum gases and (\ref{9}) reduces to the
kinetic equation found in \cite{NTL91,H90}.\\[2mm]
(v) With the medium effects on collisions, all $\Delta$'s become
independent. A generally nonparabolic momentum-dependency of the
quasiparticle energy does not allow to separate the center-of-mass
motion, and an anisotropic, inhomogeneous and time-dependent
distribution in the Pauli blocking ruins all symmetries
of the collision. The energy conservation does not reduce to the simple
conservation of the kinetic energy. The uncompensated residuum of the
energy gain contributes to a conversion between the kinetic and the
configuration energies of the system.

The kinetic equation (\ref{9}) is numerically tractable by recent Monte
Carlo codes. Kortemayer, Daffin and Bauer \cite{KDB96} have already
studied the kinetic equation with an intuitive extension by Enskog-type
displacement, claiming only a little increase in numerical demands
compared to the BE. The larger set of $\Delta$'s in (\ref{9}) does not
require principal changes of the numerical method used in \cite{KDB96}.
The functions $\varepsilon$, $z$, $|T_{\rm sc}^R|^2$, and $\Delta$'s
should form a consistent set, i.e., it is preferable to obtain them from
Green's function studies, e.g., like \cite{ARS94}. Eventual fitting
parameters should enter directly the effective nucleon-nucleon
interaction.

\section{Observables}
Let us presume that a reasonable set of functions $\varepsilon$, $z$,
$|T_{\rm sc}^R|^2$, and $\Delta$'s is known. One can then proceed to
evaluate observables. Here we present the density $n_a$ of particles
$a$, the density of energy $\cal E$, and the stress tensor
${\cal J}_{ij}$.

The observables in question are directly obtained from balance equations
which also establish conservation laws. Integrating the kinetic equation
(\ref{2}) over momentum $k$ with factors $\varepsilon_1,k,1$, one finds
that each observable has the standard quasiparticle part following from
the drift
\begin{eqnarray}
{\cal E}^{\rm qp}
&=&\sum_a\int{dk\over(2\pi)^3}{k^2\over 2m}f_1
\nonumber\\
&+&{1\over 2}\sum_{a,b}\int{dkdp\over(2\pi)^6}
T_{\rm ex}(\varepsilon_1+\varepsilon_2,k,p,0)f_1f_2,
\nonumber\\
{\cal J}_{ij}^{\rm qp}&=&\sum_a\int{dk\over(2\pi)^3}\left(k_j
{\partial\varepsilon_1\over\partial k_i}+
\delta_{ij}\varepsilon_1\right)f_1-
\delta_{ij}{\cal E}^{\rm qp},
\nonumber\\
n_a^{\rm qp}&=&\int{dk\over(2\pi)^3}f_1,
\label{10a}
\end{eqnarray}
and the $\Delta$-contribution following from the nonlocality of the
scattering integral
\begin{eqnarray}
\Delta {\cal E}&=&{1\over 2}\sum_{a,b}\int{dkdpdq\over(2\pi)^9} P
(\varepsilon_1+\varepsilon_2)\Delta_t,
\nonumber\\
\Delta {\cal J}_{ij}&=&{1\over 2}
\sum_{a,b}\int{dkdpdq\over(2\pi)^9} P
\left[(p\!+\!q)\Delta_4+(k\!-\!q)\Delta_3-p\Delta_2\right],
\nonumber\\
\Delta n_a&=&\sum_b\int{dkdpdq\over(2\pi)^9} P \Delta_t,
\label{10}
\end{eqnarray}
where $P=|T_{\rm sc}^R|^22\pi\delta(\varepsilon_1\!+\!\varepsilon_2\!-
\!\varepsilon_3\!-\!\varepsilon_4)f_1f_2(1\!-\!f_3\!-\!f_4)$. The
arguments denoted by numerical subscripts are identical to those used in
(\ref{1}), for all $\Delta$'s are explicit. The T-matrix
$T_{\rm ex}$ used in the quasiparticle part of energy is the real
part of the antisymmetrized Bethe-Goldstone T-matrix, $T^R_{\rm ex}
=(1-\delta_{ab})T^R_{\rm sc}+\delta_{ab}\sqrt{2}T^R_{\rm sc}$. The
actual observables are sum of the quasiparticle part and the
$\Delta$-correction.

In the low density limit, the $\Delta$-contributions (\ref{10}) become
proportional to the square of density. Therefore they turn into the
second order virial corrections. In the degenerated system, the density
dependence cannot be expressed in the power-law expansion since the
T-matrix, and consequently all $\Delta$'s, depend on the density.
Nevertheless, we find it instructive to call the $\Delta$-contribution
the virial corrections because of their similar structure.

The total energy ${\cal E}={\cal E}^{\rm qp}+\Delta{\cal E}$ conserves
within kinetic equation (\ref{2}). This energy conservation law
generalizes the result of Bornath, Kremp, Kraeft and Schlanges
\cite{BKKS96} restricted to non-degenerated systems. At degenerated
systems, a new mechanism of energy conversion appears due to the medium
effect on binary collisions. This mechanism can be seen writing the energy
conservation ${\partial\over\partial t}{\cal E}=
{\partial\over\partial t}{\cal E}^{\rm qp}+
{\partial\over\partial t}\Delta{\cal E}=0$ in a form
\begin{equation}
{\partial{\cal E}^{\rm qp}\over\partial t}=
\sum_a\int{dk\over(2\pi)^3}\varepsilon{\partial f_1\over\partial t}-
\sum_{a,b}\int{dkdpdq\over(2\pi)^9}P\Delta_E.
\label{12}
\end{equation}
The first term on the right hand side is the drift contribution to the
energy balance. It is the only mechanism which appears in the absence of
the virial corrections. The second term is the mean energy gain. It
provides the conversion of the interaction energy controlled by
two-particle correlations into energies of single-particle excitations.
By this mechanism, the latent heat hidden in the interaction energy is
converted into ``thermal'' excitations. The non-zero energy conversion,
$\Delta_E\not=0$, results from the time-dependency of the scattering
phase shift on the quasiparticle distribution via the Pauli blocking of
internal states. Accordingly, the energy conversion is a consequence of
the in-medium effects.

The energy gain has its space counterpart in the momentum gain to the
stress forces ${\partial\over\partial r_j}{\cal J}_{ij}$. The energy
density contributing to the stress tensor (\ref{10a}) has the gradient
\begin{equation}
{\partial{\cal E}^{\rm qp}\over\partial r_i}=
\sum_a\int{dk\over(2\pi)^3}\varepsilon{\partial f_1\over\partial r_i}-
\sum_{a,b}\int{dkdpdq\over(2\pi)^9}P\Delta_K.
\label{13}
\end{equation}
In the absence of the virial corrections (\ref{10}), the first term of
(\ref{13}) combines together with the derivative of the first term in
(\ref{10a}) into the standard quasiparticle contribution. In the
presence of the virial corrections, the energy gain is necessary to
obtain the correct momentum conservation law.

The density of energy ${\cal E}={\cal E}^{\rm qp}+\Delta{\cal E}$
given by (\ref{10a}) and (\ref{10}) alternatively results from Kadanoff
and Baym formula,
\begin{equation}
{\cal E}=\sum_a\int{dk\over(2\pi)^3}\int{d\omega\over 2\pi}
{1\over 2}\left(\omega+{k^2\over 2m}\right)G^<(\omega,k,r,t),
\label{11}
\end{equation}
with $G^<$ in the extended quasiparticle approximation (\ref{5}).
The particle density $n_a=n_a^{\rm qp}+\Delta n_a$ obtained from
(\ref{5}) via the definition, $n_a=\int{d\omega\over 2\pi}
{dk\over(2\pi)^3}G^<$, also confirms (\ref{10a}) and (\ref{10}).
The equivalence of these two alternative approaches confirms that the
extended quasiparticle approximation is thermodynamically consistent
with the nonlocal corrections to the scattering integral.

For equilibrium distributions, formulas (\ref{10a}) and (\ref{10})
provide equations of state. Two known cases are worthy of comparison.
First, the particle density $n_a=n_a^{\rm qp}+\Delta n_a$ is identical
to the quantum Beth-Uhlenbeck equation of state \cite{BKKS96,MR94},
where $n_a^{\rm qp}$ is called the free density and $\Delta n_a$ the
correlated density. Second, the virial correction to the stress tensor
has a form of the collision flux contribution known in the theory of
moderately dense gases \cite{CC90,HCB64}.

\section{Summary}
In this Letter we have derived the kinetic equation (\ref{9}) which is
consistent with thermodynamic observables (\ref{10a}) and (\ref{10}) up
to the second order virial coefficient. The presented theory extends the
theory of quantum gases \cite{NTL91,H90} and non-ideal plasma
\cite{BKKS96} to degenerated system. The most important new mechanism is
the energy conversion which follows from the medium effect on binary
collisions.

The proposed corrections can be evaluated from known in-medium
T-matrices and incorporated into existing Monte Carlo simulation codes,
e.g., with the routine used in \cite{KDB96}. The nonlocal corrections to
the scattering integral and corresponding second order virial
corrections to thermodynamic observables enlighten the link between the
kinetic equation approach and the hydrodynamical modeling. With expected
progress in fits of the single-particle energy and other ingredients of
the kinetic equation, the virial corrections can improve our ability to
infer the equation of state from the heavy ion collision data.

The authors are grateful to P. Danielewicz, D. Kremp and  G. R\"opke
for stimulating discussions.
This work was supported from Grant Agency of Czech Republic under
contracts Nos.~202960098 and 202960021, the BMBF (Germany) under
contract Nr.~06R0884, the Max-Planck-Society and the EC Human Capital and 
Mobility Programme.

\end{document}